\documentclass[letter]{aa} % for the research notes
\usepackage{graphicx}
\usepackage{txfonts}
\begin{document}
\title{Intranight variability of 3C 454.3 during its 2010 November outburst}

\author{R. Bachev\inst{1}, E. Semkov\inst{1}, A. Strigachev\inst{1}, B. Mihov\inst{1}, A. C. Gupta\inst{2}, S. Peneva\inst{1}, E. Ovcharov\inst{3}, A. Valcheva\inst{3}, A. Lalova\inst{3} }

\institute{Institute of Astronomy and National Astronomical Observatory, Bulgarian Academy of Sciences, 72 Tsarigradsko Chaussee Blvd., 1784 Sofia, Bulgaria; 
              \email{bachevr@astro.bas.bg}
         \and
             Aryabhatta Research Institute of Observational Sciences (ARIES), Manora Peak, Nainital – 263129, India\\
	   \and 
              Department of Astronomy, University of Sofia, 5 James Bourchier, 1164 Sofia, Bulgaria\\
}

\date{Received 3 February 2011; accepted 14 February 2011}

\abstract
{3C 454.3 is a very active flat spectrum radio quasar (blazar) that has undergone a recent outburst in all observed bands, including the optical.}
{In this work we explore the short-term optical variability of 3C 454.3 during its outburst by searching for time delays between different optical bands. 
Finding one would be important for understanding the evolution of the spectrum of the relativistic electrons, which generate the synchrotron jet emission.}
{We performed photometric monitoring of the object by repeating exposures in different optical bands ($BVRI$). 
Occasionally, different telescopes were used to monitor the object in the same band to verify the reliability of the smallest variations we observed.}
{Except on one occasion, where we found indications of a lag of the blue wavelengths behind the red ones, the 
results are inconclusive for most of the other cases.
There were either no structures in the light curves to be able to search for patterns, or else different approaches led to different conclusions. }
{}
\maketitle

\section{Introduction}

Blazars, perhaps the most amazing class in the AGN family, are known for their very broad-band spectra (from radio to TeV-energy gamma rays) and fast, 
large-scale variations in all bands. The commonly accepted paradigm invokes a relativistic jet, pointed almost directly toward the observer to account for 
the emission generation and the observed properties of these objects. The jet generally produces a two-peak spectral energy distribution (SED) with a lower energy synchrotron peak and a higher-energy Compton one. When signatures of an accretion disk (and a broad line region) are also present in the SED, the 
object is classified as a flat spectrum radio quasar (FSRQ).

The FSRQ 3C 454.3 (PKS 2251+158, $z=0.859$) is among the most active blazars ever observed. It is known for its violent outbursts in the optical-to-$\gamma$-ray 
bands and has been a target of numerous multiband monitoring campaigns (e.g. Villata et al. 2006; Fuhrmann et al. 2006; Raiteri et al. 2008b; 
Villata et al. 2009; Pacciani et al. 2010).  
During its latest 2010 October-November outburst it was monitored by multiple observatories in the optical (Larionov et al. 2010; Semkov et al. 2010), mm-band 
(Gurwell \& Wehrle 2010), infrared (Carrasco et al. 2010), X-, and $\gamma$-rays Vercellone et al. 2010; Stirani et al. 2010), all showing significant 
flux increase, reaching levels close to its historical maximum of 2005 May (Villata et al. 2006; see also Poggiani 2006, for a historical light curve).

Although blazar optical variability has been extensively studied throughout the years, the processes responsible for the observed, significant short-term variations 
still remain unclear. One possibility is connected with a fast evolution of the energy density distribution of the synchrotron emitting particles owing primarily to 
synchrotron/Compton losses or a fresh energy injection (acceleration). If so, there should be wavelength-dependent time lags between light curves of 
various colors, provided the optical continuum is dominated by the synchrotron emission (see the Discussion section for details). 

To the best of our knowledge only a few objects have been studied for interband time delays on the intranight timescales, and 3C 454.3 is not one of them. 
For instance, B\"{o}ttcher et al. (2010) studied the VHE blazar 1ES 1011+496 and found indications of a $\sim$5 min (but consistent with zero) lag of $B$-band 
behind the $R$-band, but considering the uncertainties and the limitations of the data-acquiring technique no specific claims were made.
During the 2006 monitoring campaign of 3C 279, based on a long-term light curve, B\"{o}ttcher et al. (2007) found a lag of about 1 and 4 days of $V$ and $B$-bands, 
respectively, behind the $R$-band, but the trend did not extend into the $I$-band. They interpreted a possible hard lag as the result of a gradual electron spectrum 
hardening due to propagation of relativistic shock front, build up of hydromagnetic turbulence, magnetic field configuration changes, etc., or alternatively, as 
a slow acceleration mechanism with a characteristic time that was close to the observed lag.

Motivated by these previous studies and considering that for many objects the results were either inconclusive or allowed various interpretations, 
we performed an intensive intranight optical ($BVRI$) monitoring of 3C 454.3 during its latest outburst, covering a total of more than 30 hours. 
Our main goals were to study the characteristics of the shortest time-scale variations of this object and, especially, to search for any possible 
time delays between the bands. This letter presents our results.

\section{Observational data}

The source 3C 454.3 was monitored most actively between 31 Oct. 2010 and 06 Nov. 2010 and mostly with two telescopes: the 50/70cm Schmidt camera (Rozhen NAO), equipped with an FLI PL16803 CCD and the 60cm Cassegrain telescope of Belogradchik AO, equipped with an FLI PL9000 CCD. Both cameras are equipped with standard 
$UBVRI$ filter sets. Occasional observations were also performed with the 2m RCC telescope of Rozhen NAO, equipped with a 
Princeton Instruments VersArray:1300B CCD. All observational frames (a total of almost 1000) were reduced, and aperture photometry (aperture radius of 4 arcsec, 
typically three times the seeing) was performed to extract the stellar magnitudes. The object was measured with respect to the nearby star "H" (Fiorucci et al., 1998), 
but other standards were also employed as check stars to verify the photometric stability. The sky was generally clear during all observational nights.
During six nights we monitored 3C 454.3 for four to six hours quasi-simultaneously in four colors ($BVRI$) in order to study the nature of the micro variability 
of the object and to search for possible time lags between the changes in different bands. Repeated $BVRI$ (only $VRI$ in one night) frame sequences 
were taken with a typical exposure time of 120 sec, thus ensuring that a frame in the same filter will be taken at least once every six to ten minutes.
Table 1 summarizes all observations.

\begin{figure}[t]
\resizebox{9cm}{!}{\includegraphics[width=9cm]{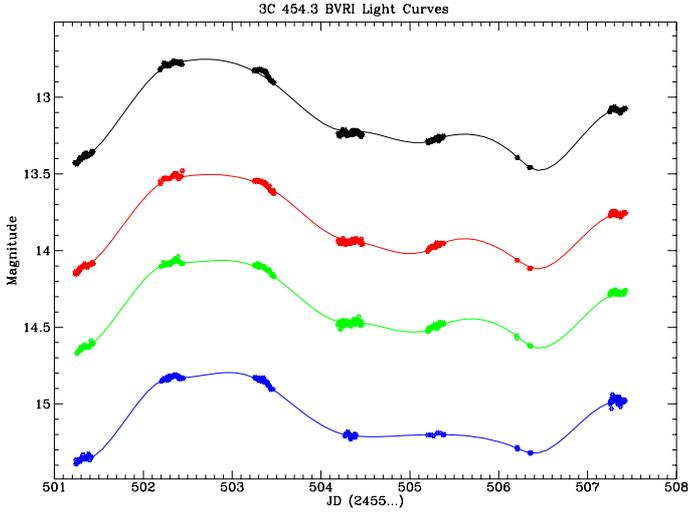}}
      \caption{$BVRI$ light curves (from bottom to top, in blue, green, red, and black, respectively) of 3C 454.3 during the most active monitoring period. 
Splines are added to guide the eye. Additional BVR data points for this observational period can be found from 
the Yale Fermi/SMARTS project (http://www.astro.yale.edu/smarts/glast/).
}
      \label{FigureLC}
\end{figure}

%\tiny
\begin{table}
\caption{3C 454.3 -- observations and measured magnitudes.}      % title of Table
\label{table:1}      % is used to refer this table in the text
\centering                          % used for centering table
\begin{tabular}{c c c c c c}        % centered columns (4 columns)
\hline\hline                 % inserts double horizontal lines
JD 2455... & Telescope\tablefootmark{a} & B & V & R & I \\    % table heading 
\hline                        % inserts single horizontal line

501.24	&	R50	&	15.38	&	14.67	&	14.15	&	13.43	 \\
501.42	&	R50	&	15.35	&	14.61	&	14.08	&	13.36	 \\
502.20	&	R50	&	14.85	&	14.10	&	13.56	&	12.82	 \\
502.42	&	R50	&	14.83	&	14.08	&	13.52	&	12.77	 \\
502.44	&	R200	&	14.83	&	14.09	&	13.48	&	12.77	 \\
503.25	&	R50	&	14.83	&	14.10	&	13.55	&	12.83	 \\
503.47	&	R50	&	14.91	&	14.17	&	13.61	&	12.91	 \\
504.20	&	B60	&	\ldots&	14.48	&	13.94	&	13.24	 \\
504.26	&	R50	&	15.20	&	14.47	&	13.93	&	13.22	 \\
504.38	&	R50	&	15.21	&	14.47	&	13.94	&	13.22	 \\
504.46	&	B60	&	\ldots&	14.48	&	13.96	&	13.24	 \\
505.20	&	B60	&	15.20	&	14.52	&	14.00	&	13.29	 \\
505.34	&	B60	&	15.19	&	14.48	&	13.96	&	13.27	 \\
505.38	&	R50	&	15.20	&	14.48	&	13.95	&	13.26	 \\
506.21	&	B60	&	15.29	&	14.56	&	14.06	&	13.39	 \\
506.35	&	R50	&	15.32	&	14.62	&	14.11	&	13.46	 \\
507.26	&	B60	&	14.98	&	14.28	&	13.77	&	13.09	 \\
507.26	&	R50	&	14.99	&	14.29	&	13.77	&	13.09	 \\
507.41	&	R50	&	14.98	&	14.28	&	13.76	&	13.08	 \\
507.42	&	B60	&	14.98	&	14.26	&	13.75	&	13.07	 \\
513.23	&	R50	&	15.01	&	14.32	&	13.80	&	13.15	 \\
538.33	&	R50	&	15.47	&	14.80	&	14.28	&	13.63	 \\
539.22	&	R50	&	15.40	&	14.69	&	14.20	&	13.53	 \\

\hline                                   %inserts single line
\end{tabular}
\tablefoot{
\tablefoottext{a}{
R50 -- Rozhen 50/70cm Schmidt telescope; R200 -- Rozhen 200cm RCC telescope; B60 -- Belogradchik 60cm telescope.} 
}
\end{table}
%\normalsize

\section{Results}
Figure 1 shows the $BVRI$ light curves of 3C 454.3 during the period when the object was most actively monitored. 
Clearly, the object was in a high state, close to its historical maximum of May, 2005 (R$\simeq$12, Villata et al. 2006) but 
still $\sim 1.5-2$ magnitudes fainter. Table 1 also gives the magnitudes at 
the beginning and the end of each sequence for the longer lasting sequences and the time-averaged magnitudes for the shorter (less than 30 min) sequences. 
We show no photometric errors in the table, but they were all in the range 0.01 -- 0.015 ($B$-band) and 0.005 -- 0.01 ($VRI$-bands) for the individual frames.

\begin{figure}
   \centering
\resizebox{9cm}{!}{\includegraphics[width=9cm]{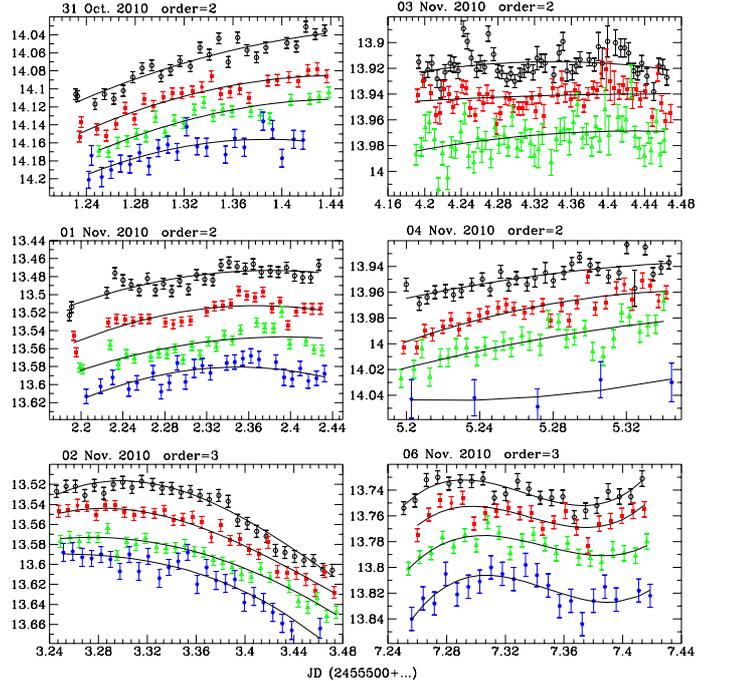}}
     \caption{Intranight $BVRI$ light curves (from bottom to top) of 3C 454.3. The evening date is shown on top of each panel, as is the
order of the polynomials used to fit the light curve (see the text). Except for $R$-band, arbitrary magnitude offsets are applied to the other 
light curves for presentation purposes.}
     \label{FigureDel}
\end{figure}

   \begin{figure}
   \centering
%\begin{minipage}[t]{0.60\linewidth}
  %\resizebox{9cm}{!}{\includegraphics[width=9cm]{3C454con.eps}}
\includegraphics[width=6cm]{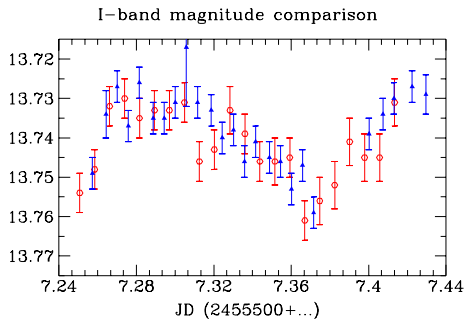}
      %\epsfig{file=3C454con.eps, width=\linewidth}

%\end{minipage}\hfill
 % \begin{minipage}[t]{0.40\linewidth}
      \caption{$I$-band light curves (with an arbitrary offset, see Fig. 2) for the night of 06 Nov. 2010 obtained at two different telescopes: 
the 60cm Belogradchik telescope (filled blue triangles) and the 50/70cm Rozhen Schmidt telescope (open red circles).}
  %  \end{minipage}
         
\label{FigureLC}
   \end{figure}

Figure 2 shows the results of the intranight monitoring of 3C 454.3 during six nights (each evening date at the top of each panel). 
For presentation purposes, the magnitudes of each band, except for $R$-band, were adjusted with a proper value in each panel of this figure 
(see Fig. 1 and Table 1 for the true $BVI$ magnitudes). To verify the reliability of these small-scale variations, the object 
was monitored with two different telescopes on some nights (Fig. 3). The nature of intranight variations of 3C 454.3 during the monitoring period can 
be described mostly as smooth, slow trends or wobbles, with no significant frame-to-frame variations and no long-term trends more than 
0.03 mag/hour (see however Raiteri et al. 2008a, for a report of much higher variation rates for another epoch of monitoring of 3C 454.3 
with an otherwise similar overall brightness). Therefore, an appropriate approach to describing the light curve in such a case is to fit a 
smooth function, such as a polynomial. In most cases our light curves could be fitted successfully with a polynomial of order 2, and occasionally 
an order 3 polynomial was used. The order of the polynomial is shown at the top of each panel in Fig. 2 and is the same for all bands during 
that night (polynomial coefficients differ from band to band, of course). Using polynomials enables us to better reveal the character of the 
variations -- the color changes, and the time lags between bands, thus suppressing the influence of the individual photometric errors and the 
absence of precise simultaneity of the frames in different bands.

A close inspection of the individual fits during a single night of observation reveals rather complex color variability behavior with no 
clear wavelength dependence from band to band (Fig. 2); i.e., the ($R-I$) color does not necessarily evolve similarly to the ($V-R$) color, etc. 
This should of course be expected if there is a time lag between the changes of different wavelengths. To search for time lags, we used the 
polynomials, fitting the light curves instead of using the light curves themselves by applying the interpolation CCF or discrete CCF techniques 
(Gaskell \& Sparke 1986; Edelson \& Krolik 1988) mainly for the following reason.
The variability amplitudes are not much larger than the photometric errors. This, combined with any possible spurious magnitude deviations affecting 
some sections of the light curve because of variable seeing, imperfect guiding combined with imperfect flat-field correction, etc. 
(see Klimek et al. 2004 for a discussion of these issues), will lead to an artificial, close-to-zero lag, based on a few deviating points at all wavelengths. 
Instead, using the fits, time lags of about an hour or less would be revealed better without allowing spurious deviations to take over the 
cross-correlation function.

Polynomial fits allow two approaches to searching for time delays. The first one is based on determining the time corresponding to a certain feature of the light 
curve (e.g. minimum or maximum) for different bands. This approach is simple and has the advantage that 
it can be applied even if the maximum (minimum) happens to be slightly outside the monitoring timeframe for some waveband. The other approach 
is based on the cross-correlation as a function of the time lag between the two polynomials; i.e., it is basically finding the maximum of the function
\begin{equation} 
CC_{xy} (\tau )=\frac{\int_{t_{\rm 1}}^{t_{\rm 2}}\left(x(t+\tau )-\bar{x}\right)\left(y(t)-\bar{y}\right)dt }{\sqrt{\int_{t_{\rm 1}}^{t_{\rm 2}}\left(x(t+\tau )-\bar{x}\right)^{2}dt\cdot \int_{t_{\rm 1}}^{t_{\rm 2}}\left(y(t)-\bar{y}\right)^{2} dt}}
\end{equation} 
where $x(t)$ and $y(t)$ are the light curves in two different wavebands.
For both approaches to be applied, one needs a polynomial of order of at least two, since the linear slope does not have features repeated 
in the next-band light curve. For that reason, although shown in Fig. 2 fit with order two polynomials, the light curves of two nights (03 and 04 Nov. 2010, 
partially due to larger photometric errors and unstable conditions) could  be fitted equally well with a linear slope and are not considered further in 
the time-delay search.

\begin{table}
\caption{Time lags (in minutes) of $BVR$ bands with respect to the $I$-band.}      % title of Table
\label{table:2}      % is used to refer this table in the text
\centering                          % used for centering table
\begin{tabular}{ c c c c }        % centered columns (4 columns)
\hline\hline                 % inserts double horizontal lines

Date	&	Band	&	Feature lags	&	CC lags	 \\
\hline                                   %inserts single line
31 Oct. 2010&	R	&	($-$97)	&	$-23\pm16$	 \\
		&	V	&	($-$90)	&	$-26\pm35$	 \\
		&	B	&	$-$185	&	$-69\pm35$	 \\
\hline                                   %inserts single line
01 Nov. 2010&	R	&	$-$19	&	$0\pm	12$ \\
		&	V	&	19	&	$12\pm 20$	 \\
		&	B	&	$-$45	&	$-16\pm 16$	 \\
\hline                                   %inserts single line
02 Nov. 2010&	R	&	$-$21	&	$4\pm 8$	 \\
		&	V	&	$-$36	&	$33\pm 14$	 \\
		&	B	&	$-$75	&	$7\pm	 13$\\
\hline                                   %inserts single line
06 Nov. 2010&	R	&	6	&	$4\pm	13$ \\
		&	V	&	19	&	$19\pm 18$	 \\
		&	B	&	23	&	$24\pm 15$ \\
\hline                                   %inserts single line
\end{tabular}
\end{table}

Table 2 shows time delays (in minutes) of $BVR$-bands, measured with respect to the $I$-band light curve. Here a positive lag means that $I$-band is 
leading the other band changes and vice versa. The first two columns are the evening date of the observation (shown also in Fig. 2) and the 
waveband to be compared with the $I$-band. The next column shows the lag measured by using a distinctive feature of the light curve (minimum or maximum). 
The number is in parentheses if this feature is found to be outside the observational period based on the polynomial fits. But of course, the results should 
be taken with caution in that case. 
For 06 Nov. 2010 data, where both the maximum and minimum are present, the average time lag is used, taking into account that both times are almost 
identical (Fig. 2, bottom right panel). The last column of Table 2 gives the corresponding lags based on the maximum of the cross-correlation function.
The errors of the CCFs are estimated by performing the same analysis on MC simulated synthetic curves (30 for each original light curve), generated taking 
the actual photometric errors into account. CCF maxima are also relatively independent of the polynomial order for that particular dataset, as 
delays change by less than 10\% when order 4 polynomial was used and about 20\% for order 5.
As evident from Table 2, for most cases there were no delays that scaled systematically with the wavelength and that were consistent regardless of 
the determination method used. 
The only exception is the 06 Nov. 2010 data, where good indications of a delay of the shorter wavelengths behind 
the longer ones can be seen, despite the relatively large CCF errors.
In addition, both methods give very consistent results. The upper panel in Fig. 4 shows the cross-correlation 
functions (for $I$-band the auto-correlation function is shown) for that date. As an additional check, the DCCF applied to the 
real data points is also shown with the corresponding errors, giving consistent results. The relation between the lag and the 
wavelength with the corresponding best linear fit is shown in the lower panel of Fig. 4.

   \begin{figure}
   \centering
      %\includegraphics[bb=32 383 306 702, width=8cm]{3C454CCn.eps}
%\includegraphics{3C454CCn.eps}
%\resizebox{9cm}{!}{\includegraphics[width=8cm]{3C454CC3.eps}}
 \includegraphics[width=6cm]{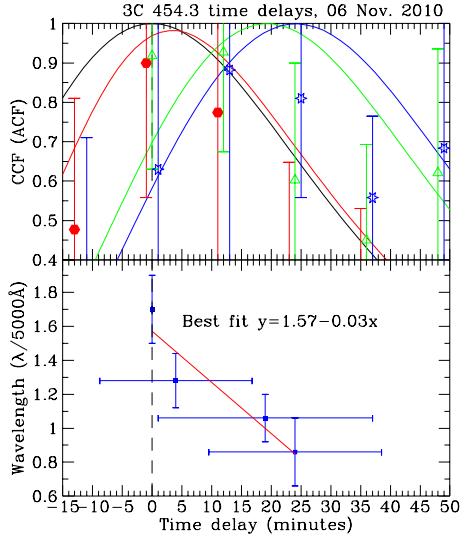}
     \caption{The cross-correlation functions used to determine the time lags during the monitoring of 06 Nov. 2010 (the upper panel). 
CCFs, applied to the fitting polynomials are in red, yellow, green and blue, green, red and black (from right to left) for 
the $BVRI$ filters, respectively (for $I$-band the auto-correlation function is shown). As an additional check, the discrete cross-correlation function 
(DCCF) applied to the real data is also shown with the corresponding errors (filled circles, open triangles, and stars for the $RVB$ bands respectively) 
and seems to indicate consistency within the errors.
The lower panel shows the relation between the wavelength of the filter (with the corresponding error based on its transparency curve) 
and the time lag with respect to the $I$-band. MC simulations are used to assess the errors of the time lags, see the text. A positive lag indicates 
a delay behind the $I$-band.
}
         \label{FigureLC}
   \end{figure}

\section{Discussion}

If the synchrotron mechanism is responsible for producing the optical emission, one would normally expect the higher frequency 
variations to precede the lower frequency ones owing to the evolution of the electron spectrum of the emitting zone. The electron energy loss 
is $d\gamma/dt\simeq -b\gamma^{2}$ for both synchrotron and IC emission (Rybicki \& Lightman 1979), where $b$ is a 
magnetic-field dependent parameter and $\gamma$ is the electron Lorentz factor. This equation indicates that the higher energy electrons, 
which are producing higher frequency photons, will evolve faster, implying that any changes in the observed optical spectrum will be observed first
in the blue wavelengths and, after some delay, in the red ones. If due to the electron energy losses, the exact time lag
can be expressed as follows (e.g. B\"{o}ttcher 2007): $\tau\simeq5(\sqrt{\lambda_{2, 5000}}-\sqrt{\lambda_{1, 5000}})$ [hours], 
where $\lambda_{1-2, 5000}$ are the two wavelengths, normalized to 5000$\AA$.
Similar wavelength-time delay behavior would also be observed in the disturbance propagating down the jet scenario (Celotti et al. 1991). 
In this model the higher energy photons are produced closer to the jet base, and a disturbance (e.g. a shock wave) gives rise to 
the corresponding frequency while traveling across the jet. 
In our $\sim30$ hours of intranight monitoring of 3C 454.3 during its recent outburst, however, we have not been able to find 
indisputable evidence for systematic red lags. One possibility is that the timescale of electron distribution evolution is longer than the typical 
monitoring time (i.e. longer than $\sim4$ hours). Another possibility implies there is an ensemble of different emitting zones, 
each evolving separately on different timescales, making it impossible to trace a color-dependent lag in the combined light curve.

On the other hand, if the variability of intranight timescales is associated with electron energy injection, one would naturally 
expect the longer wavelength variation to lead the shorter ones. 
This expectation is based on the assumption that the particle acceleration stems from some statistical process where electrons gain 
a certain amount of energy in independent, individual steps.
Thus, the higher the total energy gain of a particle, the longer the time it would take to 
reach that energy level, meaning that the energy injection process will first give rise to the longer wavelengths of the synchrotron 
spectrum.
We find strong indications but not necessarily convincing evidence for blue lags on one occasion (the night of 06 Nov. 2010) 
where the $BVR$ bands lagged consistently behind the $I$-band (Fig. 4) with the largest $\tau_{B-I}\simeq25\pm15$ min.
It is not clear whether the acceleration mechanism presumably leading to the lags observed is connected with the wave pattern of the light 
curve observed only that night. 

\section{Summary}

Our main findings can be summarized as follows.
\begin{itemize}
\item During its latest outburst, 3C 454.3 showed no rapid intranight variability (less than 0.03 mag /hour), based on about 
30 hours of monitoring.

\item Although hints of various lags between light curves of different colors were found, the overall waveband-time lag behavior 
was not systematic. Furthermore, the results were not confirmed with different methods, and so are considered inconclusive 
for most of the nights.

\item On one occasion, a wavelength-dependent lag was found with the red colors leading the blue one, perhaps an indication 
of some energy-injection process driving the variability in this case. Taking into account the errors, however, the results cannot be considered 
convincing enough. Obviously more studies are required for a firm conclusion.
\end{itemize}

\begin{acknowledgements}

The financial support from Scientific Research Fund of the Bulgarian Ministry of Education and Sciences 
through grants DO 02-85 and BIn 13/09 is acknowledged. ACG work is supported by Indo -- Bulgaria  bilateral 
scientific exchange project INT/Bulgaria/B-5/08 funded by DST, India.
Thanks are due to the anonymous referee for his/her fast report and useful suggestions.
 
\end{acknowledgements}

\end{document}